\documentclass[preprint,12pt]{elsarticle}
\usepackage{amsmath}
\usepackage{amssymb}
\usepackage{graphicx}
\usepackage{parskip}
\begin{document}

\begin{frontmatter}
%
%\shorttitle{Hyperbolic Ring Diffusion}
%\date{\today}
%\author{Peter Todd Williams}
%\shortauthor{P. T. Williams}
%\affil{The Aerospace Corporation}

%\begin{document}
%\maketitle

%\begin{abstract}

%
%\end{abstract}

%\preprint{APS/123-QED}

\title{Can Hyperbolic Diffusion Help Explain Sharp Edges in the Gaps in Saturn's Rings?}

\author{Peter Todd Williams}
\affiliation{organization={The Aerospace Corporation},%Department and Organization
            addressline={2310 E. El Segundo Blvd.}, 
            city={El Segundo},
            postcode={90245}, 
            state={CA},
            country={USA}}
%
%\affiliation{%
% The Aerospace Corporation\\% This line break forced with \textbackslash\textbackslash
% 2310 E. El Segundo Blvd. \\
% El Segundo, CA  90245
%}%


%\affiliation{
%\affiliation{
%\affiliation{%


\date{\today}% It is always \today, today,

\begin{abstract}
We explore whether hyperbolic diffusion may help explain sharp edges in the gaps in Saturn's rings.
Sharp edges are conventionally understood to be due to angular momentum flux reversal at gap edges.
We do not dispute this finding, but investigate whether non-classical diffusion may amplify this finding.
%Ordinary classical diffusion leads paradoxically to infinite speed of propagation of disturbances. In most practical
%instances, this behavior is not problematic, but there are exceptions; we suggest that planetary ring systems may be
%one example.
We explore a simple model of hyperbolic diffusion for the radial spread of material in planetary rings. The model arises by the introduction
of a relaxation time in an advection equation for the radial diffusive angular momentum flux.
%It is well-known that moons may clear
%gaps in rings on either side of the moon's orbit by resonant tidal forcing. The Encke and Keeler gaps are two canonical examples.
We show that radial secular forcing, combined with a hyperbolic diffusion equation, leads to sharp gap edges, in which the density of ring material drops
precipitously down to zero at some critical distance from the moon's orbit. Additionally, we show that our simple model can produce large
``spikes'' or ``horns'' in the density profile on either side of a ring gap, mirroring results of large N-body simulations.
It remains to be seen how these results may be affected by the inclusion of the well-understood
angular momentum flux reversal near tidally-induced gap edges.

%While we find that the relaxation time required to obtain sharp edges is unphysically large in our simple model, it is not clear 
%if this will remain the case when we allow for more physical effects, such as either a density-dependent viscosity and relaxation time,
%or an additional coupled equation for the particle velocity dispersion.
%%We leave open the question therefore whether, in the context of continuum
%%theories, hyperbolic diffusion may help explain sharp edges in ring gaps. 
%\begin{description}
%%Secondary publications and information retrieval purposes.
%\%item[Structure]
%\end{description}
\end{abstract}


%\item We propose toy model of 1D hyperbolic diffusion of planetary rings. 
%\item Ring gaps are cleared by secular tidal forcing from moons.
%\item Classical diffusion produces rounded, diffuse ring gap edges.
%\item Toy hyperbolic diffusion produces sharp edges with density enhancement on either side of gap, %qualitatively mirroring N-body results.
%\item Gap edges correspond to transition from supercritical to subcritical radial flow.


\end{frontmatter}

%\maketitle

\section{Introduction}
It is by now well-understood how at least some gaps in planetary ring systems are caused by resonant tidal forcing by moons.
In the case of Saturn, is is understood for example how the
Encke and Keeler gaps in particular
are due to overlapping high-order resonances with the co-orbital moons Pan and Daphnis respectively,
which clear out ring material on either neighboring side of their orbits.

One of the startling features of these two ring gaps is how sharp they are.
In this regard they are not unique, although they are perhaps the most noteworthy and well-studied examples.
 The surface density
of ring material does not taper off gradually as one approaches the gap; rather, the density remains more or less constant and then 
abrutly drops to near zero over a very short distance comparable to the ring thickness \cite{Long_2018}.
 On its face, this is rather surprising. 
Assuming that collisions between neighboring ring particles lead to an effective viscosity, one finds that viscous torques lead to
an effective 1D diffusion equation for ring material. Such an ordinary classical diffusion equation leads to rounded ({\em i.e.} fuzzy),
not sharp, edges \citep{Gratz_etal_2018}.

Substantial progress on this problem has been made recently by Grätz {\em et al.}~\cite{Gratz_etal_2019}, who adopt continuum methods, and by Torii {\em et al.}~\cite{Torii_etal_2024},
who adopt an N-body approach. % \cite{Torii_etal_2024}. 
% had shown previously that 
 It is true that ordinary diffusion leads to more nearly sharp edges under the assumption that viscosity is a power-law with respect to surface density, $\nu = \nu_0 (\Sigma/\Sigma_0)^\beta$, for some $\beta > 1$, 
 rather than a constant \cite{Gratz_etal_2018}. 
Indeed, N-body simulatons indicate the viscosity is proportional to the square of the the surface density
when considering self-gravity wakes\cite{DaiTanIda}.
 %\cite{Gratz_etal_2019} then
%showed how t
The inclusion of angular momentum flux reversal immediately downstream of the tidal forcing, averaged over a synodic period, results in 
even more sharp edges in a continuum theory \cite{Gratz_etal_2019}. However, even in this latter case, gap edges still have a rounded, diffusive shape, with a gradual decline in ring density before the step-function-like drop in density on either side of the gap.
Conversely, %\cite{Torii_etal_2024} %performed exhaustive 
high-fidelity N-body simulations clearly show not only
sharp gap edges, but an actual enhancement of the ring density on either side of the gap \cite{Torii_etal_2024}.

Here, following on the continuum approach, % taken by \cite{Gratz_etal_2019} and other previous researchers, 
we suggest an effect that
may be contributing to sharp edges, namely a finite relaxation time for viscous (and hence diffusive) radial flux. The inclusion of a finite relaxation time makes the parabolic diffusion equation hyperbolic, which leads to finite wave speed for the diffusive propagation of disturbances. Given a background flow (such as the secular tidally-induced radial flow), this naturally leads to a distinction between subcritial regions in which the flow speed is less than the
diffusive wave speed, and supercritical regions in which the radial flow speed is greater than the diffusive wave speed. The transition from supercritical to
subcritical flow leads to a jump condition --- akin to a shock wave or a hydraulic jump --- in which the density of ring material abruptly jumps from zero to some finite nonzero quantity.

We begin with a simple linear model for hyperbolic diffusion, with constant viscosity and constant relaxation time. This appears to substantially underpredict gap widths, assuming
conventional values for the viscosity. 
Interestingly, however,
it predicts the appearance of a large spike in density at the transition from supercritical to subcritical flow. This appears qualitatively similar to the aforementioned enhancement in the surface density profile on either side of the gap, as found by Torii {\em et al.}~\cite{Torii_etal_2024} via N-body simulation.

A simple dimensional analysis shows that hyperbolic diffusion leads to gap width scaling as $(m/M)^{1/2}$ (where $m$ is the moon's mass
and $M$ is the planet's mass), as we discuss further below. 
a simple classical viscous model
predicts $(m/M)^{2/3}$. Torii {\em et al.}~\cite{Torii_etal_2024} had found gap width 
with a possibly lower exponent, perhaps proportional to $(m/M)^{1/3}$, although this could be
due to limitations of the time limit of their simulations. Lower exponents are significant, because the higher exponent (2/3) of classical theory can not explain a number of smaller ring gaps as being due to secular tidal forcing from moons,
as the theory would predict moons massive enough that they would have been seen, in contradiction with observations~\cite{Gratz_etal_2018}.

We close by discussing possible future improvements to our model, including variable viscosity and relaxation time and a coupled energy equation for particle velocity dispersion, as well perhaps most
importantly as the inclusion of angular momentum flux reversal as discussed
by \cite{BordGoldTrem_1983}, which following the approach of \cite{Gratz_etal_2019}.

\section{Mathematical Preliminaries}
Foundational work on gap-clearing and related phenomena 
was performed by Goldreich and Tremaine \cite{GoldTrem_1980} \cite{GoldTrem_1982}, 
and Borderies, Goldreich and Tremaine \cite{BordGoldTrem_1982} \cite{BordGoldTrem_1989}, among others.

More recently, Grätz {\em et al.}~\cite{Gratz_etal_2018} studied gap formation via continuum methods, working within the Hill system of equations \cite{Hill_1878}.
They assumed that the moon mass $m$ is much smaller than the planet (Saturn) mass $M$, and that the mass of
ring particles is much less than the mass of the moon. They further assumed
that the orbit of the moon is circular, the
dynamics is two-dimensional (taking place in the plane of the rings), and the long-term effect of radial motions induced by scattering by the moon can be approximated by averaging the radial forcing over
a synodic period. The Hill system adopts a local Cartesian coordinate system, co-rotating with the moon,
with $x$ aligned with the radial direction and $y$ with the azimuthal direction. Ring particles follow 
paths that can be decomposed into an elliptical motion around a guiding center. Averaged over a synodic period, the moon induces a radial drift that can be approximated as
\begin{equation}
\frac{dx}{dt} = \frac{\alpha}{x^4} {\rm sign}({x})
\label{eq:1}
\end{equation}
with
\begin{equation}
	\alpha = \frac{{\cal A}_1^2}{18\pi} \Omega \left( \frac{m}{M} \right)^2 a_0^5
\end{equation}
mirroring previous results by Namouni~\cite{Nam_1998} and by Goldreich and Tremaine~\cite{GoldTrem_1980}.

Grätz {\em et al.}~\cite{Gratz_etal_2018} then arrived at a diffusion equation for the ring areal mass density $\Sigma$ in the
neighborhood of a perturbing moon, assuming the moon mass is large enough to clear a gap. They wrote
\begin{equation}
\partial_t \Sigma + \partial_x \left( \Sigma \frac{\alpha}{x^4} {\rm sign}(x) - 3 \partial_x (\nu \Sigma)  \right) = 0.
\label{eq:GratzDiff}
\end{equation}
In subsequent work \cite{Gratz_etal_2019}, Grätz {\em et al.} derived a modified diffusion equation,
\begin{equation}
\partial_t \Sigma + \partial_x \left( \Sigma \frac{\alpha}{x^4} {\rm sign}(x) - 3 \partial_x(\nu \Sigma K) \right) = 0
\label{eq:GratzDiff2}
\end{equation}
where $K$ parametrizes the effective shear, which undergoes reversal at the gap edge.
They found that this effect was critical in obtaining sharp gap edges.

While angular momentumm flux reversal is undoubtedly a critical effect leading to sharp gap edges,
for this preliminary work, we do not include it.
Let us then re-write eq.~(\ref{eq:GratzDiff}) as
\begin{equation}
\partial_t \Sigma + \partial_x \left( u_b \Sigma\right) = -\partial_x \Phi
\label{eq:mass}
\end{equation}
where $u_b$ is the ``background'' radial flow induced by the moon, 
{\em i.e.} $u_b = dx/dt$ from eq.~(\ref{eq:1}), and $\Phi$ is a diffusive flux
\begin{equation}
\Phi = - \partial_x \left( 3 \nu \Sigma \right).
\label{eq:flux0}
\end{equation}

Let us suppose, however, that the diffusive radial flux $\Phi$ does not respond instantaneously to changes in the 
gradient of $\nu \Sigma$. A simple model of delayed response is to indroduce a relaxation time $s$ and write
and replace eq.~(\ref{eq:flux0}), at least conceptually, with
\begin{equation}
s \dot \Phi + \Phi = - \partial_x \left( 3 \nu \Sigma \right).
\label{eq:flux1}
\end{equation}
The problem in moving from the conceptual to the concrete is how to write $\dot \Phi$ explicitly. One could use $\partial_t \Phi$, but we argue instead for the advective
derivative, choosing to write our equation for the diffusive flux $\Phi$ as
\begin{equation}
\partial_t \Phi + \partial_x \left( u_b \Phi \right) = - \frac{1}{s} \left[ \partial_x \left(3 \nu \Sigma\right) + \Phi \right]
\label{eq:flux2}
\end{equation}

The coupled pair of eqns.~(\ref{eq:mass}) and (\ref{eq:flux2}) is a modified advection-diffusion system with relaxation time. Classical diffusion is recovered in the limit $s \rightarrow 0$.
A similar set of equations arises from the modification of hyperbolic diffusion to accomodate
material advection \citep{ChrisJord_2005}. 

It is not our intent to derive eq.~(\ref{eq:flux2}) formally, but we can present a simple argument for
it based on the work of Latter \& Ogilvie \cite{LatOgi_2006}. In studying the vertically-integrated
moment equations for a dilute particulate ring, they find that the normalized visous ($xy$) stress
evolves according to (in their notation)
\begin{equation}
\partial_t \hat P_{xy} = -2 \hat P_{xy}^0 \partial_x u - P_{xx}^0 \partial_x v -\frac{1}{2}\hat P_{xx} 
+ 2 \hat P_{yy} + \hat Q_{xy}
\end{equation}
where $\hat Q_{xy}$ is a collisional term. Like Shu \& Stewart before them \cite{ShuSte_1985}, adopting
a modified BGK collision term \cite{BGK}, they find $\hat Q_{xy} \approx -\omega_c \hat P_{xy}$ where
$\omega_c$ is the collision frequency. Thus, the off-diagonal viscous stress $\hat P_{xy}$ relaxes
gradually, and does not respond instantaneously to changes in shear, behavior that is captured,
effectively, by the adoption of a relaxation term in the diffusion equation, leading to hyperbolicity.

Consider specific case $u_b=0$. Then eq.~(\ref{eq:flux2}) is a Maxwell--Cattaneo diffusion equation \citep{Maxwell_1867,Cattaneo_1948,Cattaneo_1958,Vern1958}. (But note that that moniker
 is sometimes applied to the second-order equation instead; see below.) Together,
these two equations are similar to a restricted instance of the coupled first-order telegrapher's equations, 
and can be combined into the second-order hyperbolic problem
(for $s$,~$\nu > 0$)
\begin{equation}
\partial_t \Sigma + s\, \partial_{tt}^2 \Sigma = s\, \partial_x\! \left[\frac{\partial_x(3\nu\Sigma)}{s}\right].
\end{equation}
This can be compared with the system
\begin{equation}
\partial_t \chi + s\,\partial^2_{tt} \chi = \frac{1}{c} \partial_x \left[ k \partial_x \chi\right]  
\end{equation}
where $k = k(\chi)$, a second-order equation which is also sometimes known as Maxwell--Cattaneo diffusion.

To solve the case $u_b \neq 0$ for our problem here, 
we maintain the problem as a pair of coupled first-order equations. This hyperbolic system posesses two
characteristics, $u_x \pm \sqrt{3\nu/s}$. For $|u_x| > \sqrt{3\nu/s}$, the flow is
supercritical, and no information can travel upstream, akin to shooting channel flow or supersonic
compressible flow. Then, in the steady case, this leads to a gap $\Sigma \rightarrow 0$, followed
by a hydralic jump at the critical point where $|u_x| = \sqrt{3\nu/s}$

%Now let us consider the case of the 1D radial flow of matter in Saturn's rings that is tidally induced by a %small gap-clearing moon.
%%T%he effect of the tidal forcing is approximated as a secular azimuthal forcing (torque) that induces a %radial drift $u_b$ that is a strong
%function of the radial separation $x$ between moon and ring particle guiding center. 

\section{Dimensional Analysis}

Given $\alpha$ and $\nu$, we can define a viscous lengthscale
\begin{equation}
L_v = \left( \frac{\alpha}{3 \nu} \right)^{1/3}
\end{equation}
which characterizes the width of a gap cleared by a moon, assuming ordinary classical viscosity $\nu$ is acting to re-fill the gap cleared by
the secular forcing characterized by $\alpha$.
Conversely, given a characteristic wave speed $\sqrt{3\nu/s}$, we can define a critical lengthscale
\{Note to ed: note corrected exponent.\}
\begin{equation}
L_c = \left( \frac{\alpha^2 s}{3\nu} \right)^{1/8}.
\end{equation}
When $L_v \gg L_c$, we expect the solution to approach that of classical diffusion. As $s$ increases, $L_c$ approaches $L_v$, and the solution
experiences a jump at the point of transition from supercritical to subcritical flow. The ratio $L_c/L_v$ is an extremely weak function of the dimensionless number $X$ where
\begin{equation}
X = \frac{(3\nu)^5 s^3}{\alpha^2}
\end{equation}
and $L_c/L_v = X^{1/24}$.
(Two additional physical lengthscales, $(\alpha s)^{1/5}$ and $\sqrt{3\nu s}$, can be related to $L_v$ via $X$ as well.)
When $X \gg 1$, the behavior approaches classical diffusion, but as $X$ approaches unity, and especially for $X < 1$, hyperbolic behavior
becomes more and more dominant (see below).

The gap width is set by $L_c$ as discussed below. Then since $L_c$ is proportional to $\alpha^{1/4}$, and
$\alpha$ is proportional to $(m/M)^2$, we find gap width proportional to $\sqrt{m/M}$ as mentioned previously. A weaker power-law dependence of gap width upon satellite mass is significant as, in 
principle, it allows small gaps to be cleared by proportionately smaller (less massive) moons.

%For Encke, adopting a viscosity of $250~{\rm cm~s^{-2}}$ and a relaxation time $s$ equal to the orbital %period, one finds 
%$\log_{10} X \simeq -20.5$; similarly, for Keeler, $\log_{10} X \simeq -13.5$. Correspondingly, $L_c$ is rather small compared to the gap width.
%In the case of Encke, $L_c = 22.7\ {\rm km}$, compared to a gap width of roughly $325\ {\rm km}$, and for Keeler, 
%$L_c = 3.0\ {\rm km}$, compared to a gap width of roughly $40\ {\rm km}$.

\section{Solution}
\begin{figure}
  \includegraphics[width=\columnwidth]{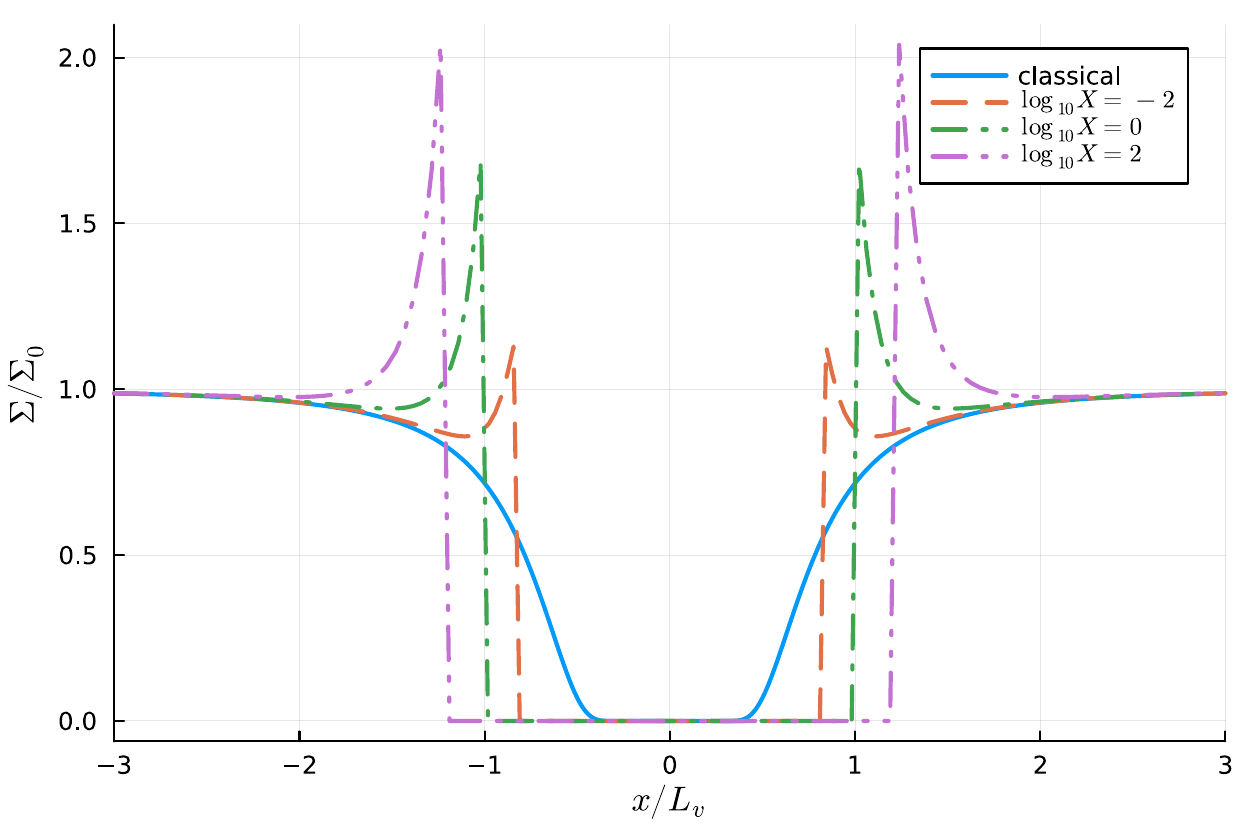}
    \caption{Solution of the toy hyperbolic diffusion equation for gap-clearing,
    for a range of values for $X$.}
    \label{fig:horns}
\end{figure}

The the coupled differential equations ({\ref{eq:mass}}) and ({\ref{eq:flux2}}) may be solved exactly
for the steady-state case $\partial_t \Sigma = \partial_t \Phi = 0$ (see fig.~\ref{fig:horns}).

Let us define $\tilde \Sigma = \Sigma L_v^2$ and $\tilde \Phi = \Phi L_v^6/\alpha$, and
let $y = x/L_v$ and define $\ell = L_c/L_v=X^{1/24}$. Then
from ({\ref{eq:mass}}) we find $\tilde \Phi = - \tilde \Sigma / y^4$, and the associated constant
of integration is zero, assuming $\tilde \Phi \rightarrow 0$ as $y \rightarrow \infty$.
Then ({\ref{eq:flux2}}) reduces to
\begin{equation}
\partial_y \left[\tilde \Sigma - \frac{\ell^8}{y^8} \tilde \Sigma \right] = \frac{1}{y^4} \tilde \Sigma,
\end{equation}
with solution
\begin{equation}
\tilde \Sigma = \frac{y^8 \tilde \Sigma_0 }{y^8 - \ell^8} \exp{\left( \frac{-f(|y|/\ell)}{ \ell^3}\right)}
\end{equation}
when $|y|>\ell$, and $\tilde \Sigma = 0$ otherwise, where
\begin{multline}
f(z) = \int_z^\infty \frac{dz'}{z'^4 -1/z'^4}  = 
\frac{1}{16} \left[ 2\pi\left( i-1+\sqrt{2} \right) + 4 \tan^{-1}(z) + \right. \\
\left. + 4 \tanh^{-1}(z) + \sqrt{2}\left( 2 \tan^{-1}(1-\sqrt{2} z) - 2\tan^{-1}(1+\sqrt{2}z) + \right. \right. \\
\left. \left. + \log(1-\sqrt{2}z+z^2) - \log(1+\sqrt{2}z+z^2) \right) \right] 
\end{multline}
Note that in the limit $\ell \rightarrow 0$, for $y>0$ we recover the classical diffusive solution
%\cite{Gratz_etal_2018}
\begin{equation}
\tilde \Sigma = \tilde \Sigma_0 \exp\left(-\frac{1}{3y^3} \right)
\end{equation}
%We find an approximate analytic solution to the problem by finding the first
%twenty terms in an asymptotic series (including only negative or zero powers) for
%$\Sigma$ and $\Phi$.
%and corroborating are confirmed a simple Godunov method \citep{Godunov_1959, Godunov_1962, LeVeque}.
That is, for $X \ll 1$, the solution approaches the diffusive solution for $\beta=1$ that was shown by
Grätz {\em et al.}~\cite{Gratz_etal_2018}. 

For $X>0$, the solution diverges weakly at a cusp at the critical point $y=\ell$, {\em i.e.} at $x=L_c$.
As $X$ increases, the cusp becomes wider and grows, becoming pronounced for $X\simeq 1$.
%For $X=1000$, the cusp grows to approximately three times the 
%height of the background density $\Sigma_0$.

\begin{figure}
  \includegraphics[width=\columnwidth]{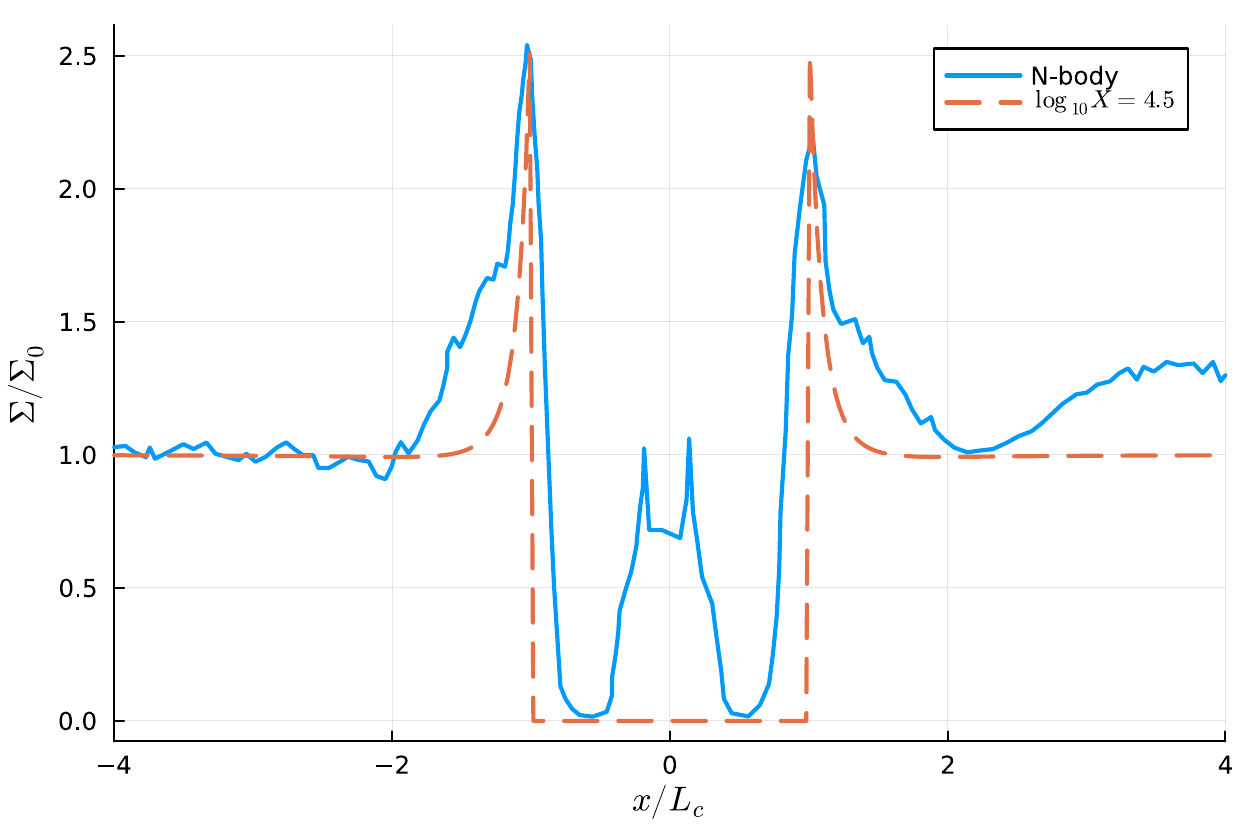}
    \caption{Comparison between toy hyperbolic problem and N-body results of Torii {\em et al.}~(2024).}
    \label{fig:Plotfit}
\end{figure}

It is interesting to compare these results with figure 8 of Torii {\em et al.}~\cite{Torii_etal_2024}, which shows qualitatively similar cusps or ``horns''
in the density profile on either side of the gaps cleared in their N-body simulations (fig.~\ref{fig:Plotfit}). As noted by
Torii {\em et al.}~\citet{Torii_etal_2024}, such features had
been seen in earlier N-body simulations \cite{LewSte_2000, Lew_etal_2011}, and appear to correpond to an observed enhancement
in optical depth on either side of ring gaps \cite{Show_etal_1986}.

On the other hand, to achieve such ``horns'' in the very simple model we have adopted requires large values of the relaxation time $s$. 
For example, for $X=1$, for Keeler, we require a relaxation time $s$ of order 50 yr, and for Encke, as much as 10,000~yr. Physically,
one expects the relaxation time to be at most a factor of few times the orbital period of $~14$~hr.

Alternatively, one may match the width of the Keeler and Encke gaps by adopting extremely low values of the
viscosity. For example, setting the relaxation time to be the orbital time scale, we can match the
Keeler gap width by setting the kinetmatic viscosity $\nu$ equal to $3.5\times 10^{-5} {\rm cm^2/s}$, and the
Encke gap with $\nu = 4.6\times 10^{-5} {\rm cm^2/s}$. These values are substantially less than
the more generally accepted values typically of order $100\ {\rm cm^2/s}$ \cite{Taj_etal_2017}.

Gap width in all cases is $L_c$, since flow is supercritical for $|x|<L_c$ yielding no steady solution
for $\Sigma > 0$ in this region. 

\section{Commentary}
We have shown that a simple hyperbolic diffusion equation can qualitatively reproduce sharp gap edges as well as ``horns'' in the ring 
density profile. On the other hand, our model was rather {\em ad-hoc}, did not address energy balance (ring particle velocity dispersion), and
required unphysically large values of the relaxation time in order to reproduce sharp gap edges. 
Most importantly, our model did not include the well-known angular flux luminosity reversal, which is
understood to contribute to sharp gap edges.
%Furthermore, basic dimensional analysis suggests
%a critical lengthscale $L_c$ that is roughly an order of magnitude smaller than the observed gap widths of %Encke and Keeler.

Despite these deficiencies, we suggest that it may be worth exploring whether hyperbolicity of the effective continuum diffusion of ring particles
may help explain sharp gap edges.

First, in general, we should expect $\nu$ and $s$ to depend upon $\Sigma$, which we did not treat in our simple model. Indeed it was shown by
Grätz {\em et al.}~\cite{Gratz_etal_2018} that even in the case of simple classical diffusion, the edges of a gap become sharper if the viscosity is assumed to scale 
proportionally to $(\Sigma/\Sigma_0)^\beta$ for some $\Sigma_0$ and some $\beta>0$, which seems reasonable.
Additionally, in our model, we might also
assume the relaxation time $s$ to behave similarly, say $s = s_0 (\Sigma/\Sigma_0)^{-\gamma}$ for some
$s_0$ and $\gamma>0$. Unfortunately, these modifications, however, make the problem nonlinear and less tractable, which is why we did not include them here.
There has been recent progress on numerical solution of hyperbolic diffusion \citep{Gomez_etal_2010, Reverberi_etal_2008}; it remains to be seen
if these methods could be applied to our more general problem.

Perhaps equally significantly, we have not addressed the corresponding advection-diffusion equation for the ring energy (or velocity dispersion).
It is not at all obvious how the inclusion of this important physical effect might affect our results.
And, as has been mentioned previously, angular momentum flux-reversal can and should be included in
our effective diffusion equation \cite{Gratz_etal_2019}.

Finally, it is worth asking whether, following the work of Ostoja-Starzewski~\cite{Ostoja-Starzewski_2009} and Zhang and Ostoja-Starzewski~\cite{Zhang_Ostoja-Starzewski_2019} among others,
 a hyperbolic advection-diffusion system could be arrived at with some degree of formalism rather than
by simple fiat. 

Awaiting this work, we therefore hypothesize in the mean time that some form of hyperbolic diffusion might help explain sharp ring gaps, and in particular, the ``horns'' in density seen in N-body simulations.

%\begin{acknowledgments}
\section*{Acknowledgments}
The author thanks Martin Ostoja-Starzewski for  introducing him to hyperbolic diffusion, and
Henrik Latter for many fruitful discussions on planetary ring physics.
%\end{acknowledgments}

%}
%\end{itemize}

% REFERENCES

\bibliographystyle{mnras}
\bibliography{razor_rev1_}{}

\end{document}